\documentclass[journal=nalefd,manuscript=letter,layout=twocolumn]{achemso}

\usepackage[version=3]{mhchem} % Formula subscripts using \ce{}
\usepackage[T1]{fontenc}       % Use modern font encodings
\usepackage[utf8]{inputenc}
\usepackage[english]{babel}
\usepackage{amsmath,amsfonts,amsthm,amssymb}
\usepackage{mathtools}
\usepackage{graphicx}
\usepackage{xcolor}
\usepackage[pdftex]{hyperref}
\usepackage{bm}
\usepackage[inline]{enumitem}
\usepackage{array}
\usepackage{gensymb}
\usepackage[font=footnotesize,labelfont=default]{caption}
\usepackage{verbatim}
\usepackage{mathrsfs}

\newcommand{\ep}{\epsilon}%dielectric function
\newcommand{\eps}{\mathscr E}%single-electron energy

\renewcommand{\vec}[1]{\bm{#1}}

\newcommand{\rrr}{\vec{r}}

\newcommand{\GGn}{\vec{G}}

\newcommand{\ggn}{\vec{g}}

\newcommand{\HHH}{\vec{H}}

\newcommand{\VVn}{\vec{V}}
\newcommand{\RRR}{\vec{R}}

\DeclareMathOperator{\re}{Re}
\DeclareMathOperator{\im}{Im}

\newcommand{\eqlab}[1]{\label{eq:#1}}
\renewcommand{\eqref}[1]{Eq.~(\ref{eq:#1})}
\newcommand{\eqsref}[2]{Eqs.~(\ref{eq:#1}) and~(\ref{eq:#2})}
\newcommand{\figref}[1]{Fig.~\ref{fig:#1}}

\author{Mikkel Settnes}
\affiliation[]{Department of Photonics Engineering, Technical University of Denmark, DK-2800~Kongens~Lyngby, Denmark}
\alsoaffiliation[]{Department of Micro and Nanotechnology, Technical University of Denmark, DK-2800~Kongens~Lyngby, Denmark}
\alsoaffiliation[]{Center for Nanostructured Graphene, Technical University of Denmark, DK-2800~Kongens~Lyngby, Denmark}
\author{J.~R.~M. Saavedra}
\affiliation[]{ICFO-Institut de Ciencies Fotoniques, The Barcelona Institute of Science and Technology, 08860~Castelldefels~ 
(Barcelona), Spain}
\author{Kristian~S.~Thygesen}
\affiliation[]{Center for Atomic-Scale Materials Design (CAMD), Department of Physics, Technical~University~of~Denmark, DK-2800~Kongens~Lyngby, Denmark}
\alsoaffiliation[]{Center for Nanostructured Graphene, Technical University of Denmark, DK-2800~Kongens~Lyngby, Denmark}
\author{Antti-Pekka~Jauho}
\affiliation[]{Department of Micro and Nanotechnology, Technical University of Denmark, DK-2800~Kongens~Lyngby, Denmark}
\alsoaffiliation[]{Center for Nanostructured Graphene, Technical University of Denmark, DK-2800~Kongens~Lyngby, Denmark}
\author{F.~Javier~Garc\'ia~de~Abajo}
\affiliation[]{ICFO-Institut de Ciencies Fotoniques, The Barcelona Institute of Science and Technology, 08860~Castelldefels~ 
(Barcelona), Spain}
\alsoaffiliation[]{ICREA-Institucio Catalana de Recerca i Estudis Avancats, Passeig Llus Companys, 23,
08010~Barcelona, Spain}
\author{N.~Asger~Mortensen}
\affiliation[]{Center for Nano Optics, University of Southern Denmark, Campusvej 55, DK-5230~Odense~M, Denmark}
\alsoaffiliation[]{Danish Institute for Advanced Study, University of Southern Denmark, Campusvej 55, DK-5230~Odense~M, Denmark}
\alsoaffiliation[]{Center for Nanostructured Graphene, Technical University of Denmark, DK-2800~Kongens~Lyngby, Denmark}

\email{asger@mailaps.org}

\title[]{Strong plasmon-phonon splitting and hybridization in 2D materials revealed through a self-energy approach}


\begin{document}

\begin{abstract}
We reveal new aspects of the interaction between plasmons and phonons in 2D materials that go beyond a mere shift and increase in plasmon width due to coupling to either intrinsic vibrational modes of the material or phonons in a supporting substrate. More precisely, we predict strong plasmon splitting due to this coupling, resulting in a characteristic avoided crossing scheme. We base our results on a computationally efficient approach consisting in including many-body interactions through the electron self-energy. We specify this formalism for a description of plasmons based upon a tight-binding electron Hamiltonian combined with the random-phase approximation. This approach is accurate provided vertex corrections can be neglected, as is is the case in conventional plasmon-supporting metals and Dirac-fermion systems. We illustrate our method by evaluating plasmonic spectra of doped graphene nanotriangles with varied size, where we predict remarkable peak splittings and other radical modifications in the spectra due to plasmons interactions with intrinsic optical phonons. Our method is equally applicable to other 2D materials and provides a simple approach for investigating coupling of plasmons to phonons, excitons, and other excitations in hybrid thin nanostructures.
\end{abstract}

\small

\bigskip

\section{Introduction}

Two-dimensional (2D) materials are receiving an increasing interest in nanophotonics due to their ability to host a large variety of polaritons,\cite{Basov:2016,Low:2017} such as tunable plasmon-polaritons in graphene\cite{Fei:2012,Chen:2012} or phonon-polaritons in hexagonal boron nitride (h-BN).\cite{Dai:2014,Caldwell:2014,Caldwell:2015} In this context, plasmonics benefits from the unique charge-transport properties of graphene,\cite{Abajo:2014,Xiao:2016,Goncalves:2016,Huang:2017} because the electron mobility in a suspended single-atomic-layer can reach 200,000\,cm$^2$V$^{-1}$s$^{-1}$.\cite{Bolotin:2008}  
Indeed, effects associated with phonon scattering in polarizable substrates\cite{Fratini:2008,Politano_2017} have been found to be one of the factors limiting the carrier mobility in graphene.\cite{Hess:1979}
Hexagonal boron nitride phonons couple weakly to graphene electrons, and is therefore suitable for graphene encapsulation compatible with preservation of high mobility.\cite{Dean:2010,Woessner:2015,Principi:2014} When the characteristic energy of plasmons coincides with that of phonons, we anticipate that plasmon-phonon hybridization can take place.\cite{Hwang:2010,Jablan:2011} Actually, strong phonon-plasmon coupled modes have been observed in graphene/SiC\cite{Liu:2010,Koch:2010} and graphene/SiO$_2$ interfaces.\cite{Fei:2011,Yan:2013,Zhu:2014} Likewise, coupling between plasmons and surface phonons has been examined in thin polar substrates.\cite{Li2014,Victor2014,Barcelos2015} Additionally, carrier mobility depends on the electrostatic environment, and has been studied in the context of electrostatic gating of graphene.\cite{Gunst:2017}
The influence of substrate-hosted phonons is commonly included through the substrate dielectric function,\cite{Abajo:2014} which qualitatively explains experimental observations rather well.\cite{Yan:2013,Zhu:2014} Unfortunately, only limited progress has been reported in the quantum description of plasmon-phonon coupling, essentially restricted to analyses based on tight-binding (TB) models combined with the random-phase approximation (RPA).\cite{Thongrattanasiri:2012a,Christensen:2014,Wang:2015,Stauber:2014} More generally, quantum-plasmonic phenomena\cite{Tame:2013,Bozhevolnyi:2017a,Bozhevolnyi:2017b} are commonly treated at the single-particle level using various mean-field models,\cite{Raza:2015a,Varas:2016} without more explicit accounts for many-body interactions.
In this manuscript, we develop an RPA-inspired formalism for the quantum plasmonic response in which the interactions are encoded in the single-electron Green functions through electron-phonon self-energies. As an illustration of our theory, we evaluate the influence of phonons on plasmon resonances in graphene nano-triangles.

Our method is amenable application to arbitrary graphene geometries and can be directly extended to other plasmon-supporting 2D materials such as black phosphorous\cite{HMP17} and thin noble metals.\cite{paper236}

\section{Methods}
Within the RPA, the dielectric matrix, $\ep(\rrr,\rrr',\omega)$, can be expressed in terms of the non-interacting polarizability $\chi^0(\rrr,\rrr'\omega)$ 
\begin{align}
\ep(\rrr,\rrr',\omega) &= 1- \sum_{\RRR} V(\rrr-\RRR) \chi^0(\RRR,\rrr',\omega). \eqlab{eps}
\end{align}
Here, $V(\rrr-\rrr') \propto e^2/|\rrr-\rrr'|$ is the bare Coulomb interaction and the summation is over the atom positions $\RRR$. The formal on-site divergence ($\rrr\rightarrow \rrr'$) is only an apparent issue because of the orbitals' finite extension, which is incorporated through a self-interaction term of 0.58 atomic units for $\rrr=\rrr'$.\cite{Thongrattanasiri:2012a,Wang:2015}
The non-interacting polarizability is 
\begin{align}
\chi^0(\rrr,\rrr',\omega)  = 2\sum_{nm} \big[f(\eps_m)-f(\eps_n)\big] \nonumber \\
 \times  \frac{\psi_n(\rrr) \psi_n^*(\rrr') \psi_m^*(\rrr)\psi_m(\rrr')}{\hbar\omega + i \eta - \eps_n +\eps_m}, \eqlab{chi0}
\end{align}
where $f(\eps)$ is the Fermi--Dirac distribution function and $\eps_n$ is the energy associated with the single-electron wavefunction $\psi_n$, while $\eta$ is an infinitesimal broadening. The leading factor of 2 originates in spin degeneracy.

Self-sustained charge density oscillations (plasmons) can exist where the dielectric matrix has zero determinant and the related potential, $\phi$, satisfies
\begin{align}
\sum_{\RRR} \ep(\rrr,\RRR,\omega)\phi(\RRR,\omega) = 0.\eqlab{Re0}
\end{align}
The dielectric matrix may have a finite imaginary part, so \eqref{Re0} cannot be satisfied in general for real $\omega$. Instead, we require that the real part vanishes, which defines the plasmon modes $\omega_n$, and the associated potential $\phi_n$ via
\begin{align}
\sum_{\RRR} \ep(\rrr,\RRR,\omega_n)\phi_n(\RRR,\omega_n) = i \ep_n \phi_n(\rrr,\omega_n).\eqlab{phi_eig}
\end{align}
Here, $\ep_n$ is the imaginary part of the eigenvalue of the dielectric matrix $\ep(\omega_n)$. Therefore, the plasmon modes are the eigenvectors corresponding to purely imaginary eigenvalues of the dielectric matrix.\cite{Andersen2012_PRB,Andersen2013_PRB}

Plasmonic spectra can be conveniently analyzed in terms of the electron- energy--loss  function $-\im \{ \ep^{-1}(\omega)\}$, which is relevant for the probing of plasmons in electron-energy loss spectroscopy.\cite{Abajo:2010} Considering the eigenvalues $\ep_n$, we therefore define the plasmon frequency, $\omega_n$, as the local maximum of $-\im \{ \ep_n^{-1}(\omega_n)\}$. For a system composed of $N$ atomic sites there exist $N$ eigenvalues and corresponding eigenfrequencies. Below, the main focus is placed on the two eigenvalues with the largest value of $-\im \{ \ep_n^{-1}(\omega)\}$, corresponding to doubly-degenerate dipolar plasmon resonances.\cite{Wang:2015} In brief, we can calculate the eigenvalue loss-spectrum and corresponding plasmon modes using the eigenvalues and eigenvectors of the dielectric matrix~\eqref{eps}. Evidently, this requires an efficient way of calculating $\chi^0(\omega)$. As we now show below, interactions with phonons can readily be included in this scheme.

It is convenient to express the polarizability in terms of the retarded Green function $G^r_0(\rrr,\rrr',\eps)$ and the spectral function $A_0(\rrr,\rrr',\eps)$:
\begin{align}
G^{r}_0(\rrr,\rrr',\eps) &= \sum_n \frac{ \psi_n(\rrr)  \psi^*_n(\rrr')}{\eps-\eps_n +i \eta},\eqlab{GFdef1}\\
A_0(\rrr,\rrr',\eps)&= i \big[G^r_0(\rrr,\rrr',\eps)-G^a_0(\rrr,\rrr',\eps) \big] \nonumber \\
&= 2\pi\sum_n \psi_n(\rrr) \psi^*_n(\rrr') \delta(\eps-\eps_n), \eqlab{GFdef2}
\end{align}
where we also use the advanced Green function, which under the assumption of time-reversal symmetry reads $G^a_0({\bf r},{\bf r}^\prime,\mathcal{E})=\left[G^r_0({\bf r}^\prime,{\bf r},\mathcal{E})\right]^*$. Using \eqsref{GFdef1}{GFdef2} while exploiting the symmetry $A(\rrr,\rrr',\eps)=A(\rrr',\rrr,\eps)$, we can express \eqref{chi0} in a more general form without any reference to single-particle wavefunctions
\begin{align}
\chi^0(\rrr,\rrr',\omega) &= \frac{1}{\pi} \int d\eps\, f(\eps) A_0(\rrr,\rrr',\eps)  \nonumber \\
\times \bigg[ G^r_0(\rrr,\rrr'&,\eps+\hbar\omega)
+ G^a_0(\rrr',\rrr,\eps-\hbar\omega)\bigg].
\end{align}
We focus on the imaginary part of $\chi^0(\omega)=\chi^0_R(\omega) +  i \chi^0_I(\omega)$. 
In the low temperature limit, we thus have\cite{Prange2009_PRB}
\begin{align}
\eqlab{chi0Im}
&\chi^0_I(\rrr,\rrr',\omega)= \\
& -  \frac{1}{2\pi}\int^{\eps_F}_{\eps_F-\hbar\omega}d{\eps}\, A_0(\rrr,\rrr',\eps) A_0(\rrr,\rrr',\eps+\hbar\omega) ,
\nonumber
\end{align}
where $\eps_F$ is the Fermi energy. We note that \eqref{chi0Im} can be conveniently written in a convolution form, which allows us to carry out an efficient numerical implementation. The real part of $\chi^0(\omega)$ can then be calculated using the Kramers--Kronig relations with an 8\,eV cut-off, as shown in the Appendix.

For any system with time-reversal symmetry, the spectral function corresponds to the imaginary part of the Green function $A(\rrr,\rrr',\eps) = -2\im\{ G(\rrr,\rrr',\eps)\}$. Consequently, \eqref{chi0Im} only requires the Green function of the system. This implies an interesting possibility to estimate the effect of interactions: the noninteracting spectral functions in \eqref{chi0Im} can be replaced by interacting Green functions with self-energy insertions.  Notably, an energy-dependent, but spatially uniform, self-energy term does not essentially increase the computational burden. The Green function including an energy dependent retarded self-energy term $\Sigma(\eps)$ can be symbolically written as (we suppress the spatial dependence here)
\begin{align}
\GGn^r(\eps) = \big[\eps + i\eta - \HHH - \Sigma(\eps)\big]^{-1}. \eqlab{GF}
\end{align}
The term $\Sigma(\eps)$  accounts for the interactions between electrons with other degrees of freedom,\cite{Haug:2008} such as phonons, but it can also account for the life-time broadening of electron states in open quantum systems where electrons can leak into semi-infinite surroundings.\cite{Datta:1995} Note that \eqref{GF} contains self-energy corrections to the one-point Green function, while a full discussion would require the analysis of a two-point function (i.e., the interacting polarizability). Thus, using \eqref{GF} (and the spectral functions associated with it) implies the omission of vertex corrections. For metals this is a common approach justified by Migdal's theorem,\cite{Bruus:2004} while similar considerations can be made for Dirac materials,\cite{Roy:2014} given that the sound velocity is much smaller than the Fermi velocity ($v_F\sim 10^6$\,m/s for graphene).  A quantitative assessment of the role of the vertex corrections is however beyond the phenomenological approach adopted in this work.

Many different approaches can be used to determine the Green function $G(\rrr,\rrr',\eps)$, including brute force inversion or analytical methods. Below, we focus on recursive techniques, which allow us to efficiently deal with large, spatially inhomogeneous systems described by a generic TB Hamiltonian.

\subsection{Recursive Green function}
We employ the recently developed efficient recursive Green function approach to obtain  $\GGn^r_0(\rrr,\rrr',\eps)$.\cite{Settnes2015_PRB,Lewenkopf2013} Dividing the system into $N_{\rm cell}$ cells (i.e., with an average of $M=N/N_{\rm cell}$ atoms each) only connecting to neighboring cells, the forward recursion is given by
\begin{subequations}
\begin{align}
\ggn_1 &= (\eps+ i\eta - \HHH_1)^{-1}, \\
\ggn_n &= (\eps + i\eta - \HHH_n - \Sigma_n)^{-1},\\
\Sigma_n & = \VVn_{n,n-1} \ggn_{n-1} \VVn_{n-1,n},
\end{align}
\end{subequations}
where $\VVn_{n,n-1}$ is the coupling matrix between cell $n$ and cell $n-1$.
The last cell contains the full Green function $\GGn$ of that cell. To obtain the entire Green function matrix, we save $\ggn_n$ from the forward recursion and do a backwards recursive sweep consisting in updating all diagonal and off-diagonal blocks according to,
\begin{subequations}
\begin{align}
\GGn_n &=  \ggn_n + \ggn_n \VVn_{n,n-1} \GGn_{n-1} \VVn_{n-1,n} \ggn_n, \\
\GGn_{n,m} &= \ggn_n \VVn_{n,n-1} \GGn_{n-1,m} .
\end{align}
\end{subequations}
This yields the total Green function for a given energy.

Using this recursive scheme, the computational scaling changes from $O(N_\omega N^3)$ to $O(N_\omega N_{\rm cell} M^3)$, where $N_\omega$ is the number of energy points required to perform the integration in \eqref{chi0Im}.

The convolution yielding $\chi^0_I(\omega)$ can be done using fast Fourier transforms, yielding a scaling of $O(N \log N)$. A similar scaling is obtained for the Kramers--Kronig transformation, giving a total scaling of the method $O(N_{\omega}N_{\rm cell}M^3)$.

\section{Plasmons in graphene nanotriangles}
We describe the electronic structure of graphene  using a nearest-neighbor TB model $H = -t_0 \sum_{\langle i,j \rangle} c_i^\dagger c_j$, where $t_0=2.8$\,eV, while the sum runs over all neighbor pairs $\langle i,j\rangle$. To account for a finite carrier density, we choose a Fermi energy $\eps_F=0.4$\,eV, which is a typical value in experiments exploring plasmon-phonon coupling.\cite{Zhu:2014} As an illustrative example, we consider graphene nanotriangles with armchair edges because these have a smooth evolution of the electronic states when increasing size.\cite{Nakada:1996,Potasz:2010} 
Graphene nanostructures such as nanodisks do not necessarily conserve their edge-configuration of atoms while varying their structure sizes (i.e. the emergence of localized edge states may drastically change both the electronic structure and the plasmon response when changing size\cite{Thongrattanasiri:2012a,Christensen:2014}). In the following, we focus on the armchair edge configuration to suppress the complexity added by the appearance of edge states formed at zigzag edges.\cite{Manjavacas:2013,Christensen:2014}

\begin{figure}[htb!]
 	\begin{center}
 		\includegraphics[width= 0.95\columnwidth]{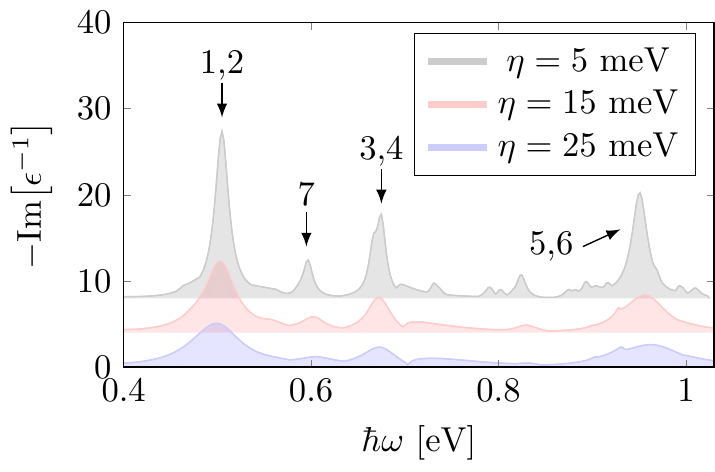}
 		\caption[]{Eigenvalue loss spectrum for an armchair nanotriangle with side length of 8\,nm and different values of the broadening $\eta$. The different modes are labeled according to \figref{eels_mode}.} \label{fig:eels_no_phonon}
 	\end{center}
 \end{figure}

\begin{figure}[htb!]
 	\begin{center}
 		\includegraphics[width= 0.95\columnwidth]{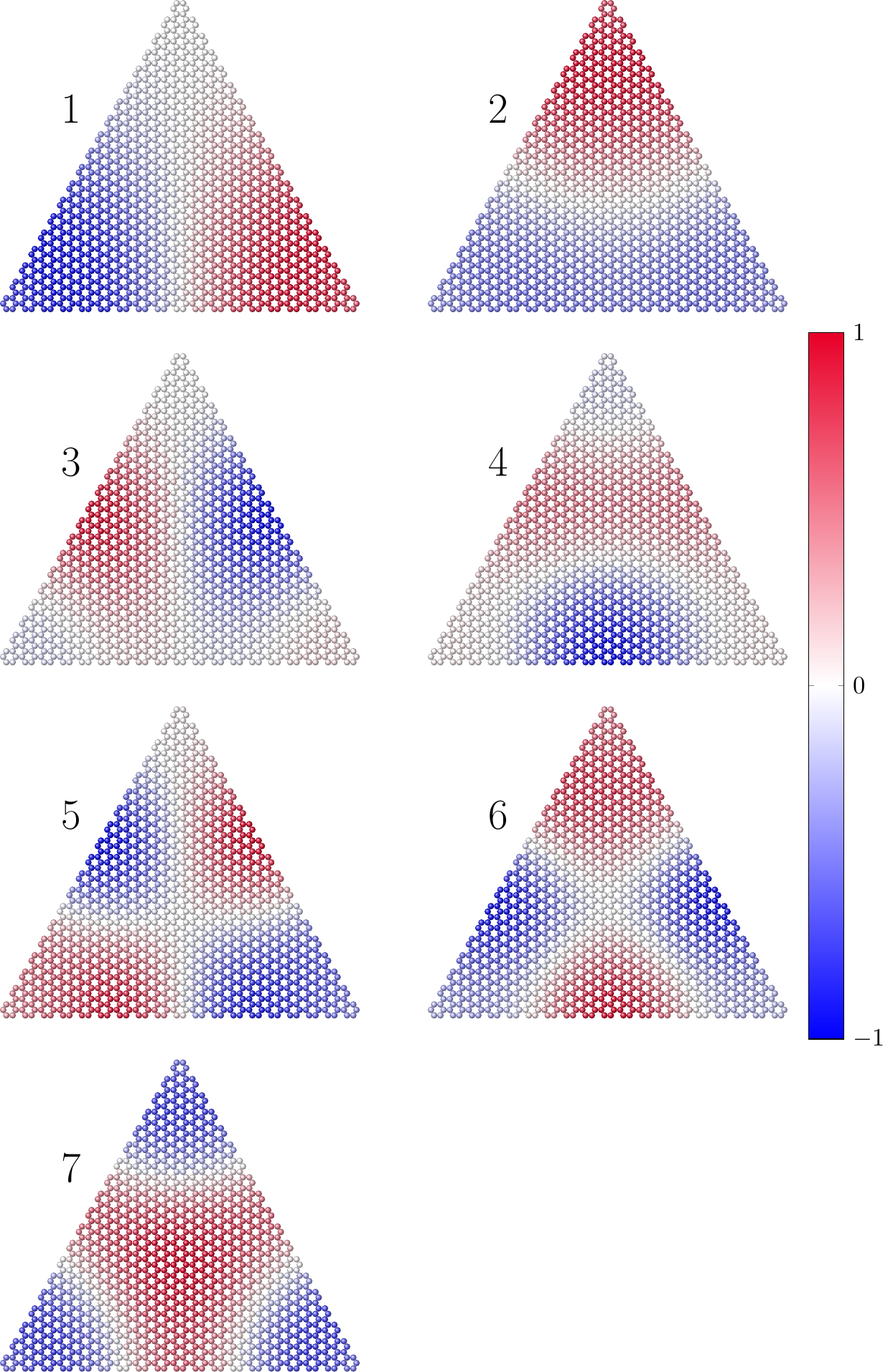}
 		\caption[]{Real space plot of the plasmon scalar potential $\re \{\phi_n\}$ for the different plasmon modes identified in \figref{eels_no_phonon}. } \label{fig:eels_mode}
 	\end{center}
 \end{figure}

First, we calculate the eigenvalue loss-spectrum for an armchair nanotriangle with side length $\sim 8.1$\,nm (see \figref{eels_no_phonon}). Multiple plasmon peaks are clearly identifiable in the spectrum.
Inspection of the individual eigenvalues reveals that each peak is either non-degenerate or consists of a pair of eigenstates with double degeneracy. This is consistent with group theory considerations.\cite{Wang:2015,Awada2012_JPC} \figref{eels_mode} shows the real part of the scalar potential eigenstate $\phi_n$ for the plasmon peaks labeled in \figref{eels_no_phonon}. Clearly, $\re \{\phi_n\}$ reveals a double degeneracy for the modes 1--6, whereas the plasmon mode labeled 7 is non-degenerate. The doubly degenerate modes 1--6 are either symmetric or anti-symmetric with respect to the mirror plane. We expect the strongest coupling  from an external optical source to the dipole modes 1--2 with electrical fields polarized along the spatial profile of the mode.

\subsection{Plasmon-phonon coupling}
Interaction with phonon modes leads to broadening of a plasmonic peaks because of the associated damping, that is captured by $\im \{\Sigma\}\ne 0$. A simple, phenomenological way to study the effects due to damping is to vary the $\eta$ parameter in \eqref{chi0}. This is been done in \figref{eels_no_phonon}, where the eigenvalue loss-spectrum is shown for different values of the broadening $\eta$. It is clear that the peaks broaden and can develop fine structure as discussed in Ref.~\citenum{Thongrattanasiri:2012a}. However, by construction $\re \{\Sigma\} =0$ in this model and there are no changes in the central positions of the spectral components.

Generally, interaction effects are energy-dependent and consequently cannot be captured accurately by a constant broadening. The effect of energy- dependent interactions can be described  using a self-energy formalism.\cite{Haug:2008} As explained above, within the present approach the different interaction effects do not increase the size of the Hilbert space, thus resulting in just a minor increase in computational demand.

In the calculations presented below, we consider a simple model for the electron-phonon interaction appropriate to describe optical phonons in graphene.
We follow the approach of Refs.~\citenum{Park2007,Carbotte2010_PRB,Jablan2009_PRB} and consider a dispersionless optical phonon of energy $\hbar\Omega_0 = 0.2$\,eV. The treatment, however, is general and can be equally applied to both intrinsic and substrate phonons. For simplicity, we assume a constant electron-phonon matrix element $g_0$ yielding the on-site self-energy in the Born approximation,\cite{Haug:2008,Park2007,Carbotte2010_PRB}
\begin{align}
&\im \{\Sigma_{\rm ph}(\eps,T)\}  =-\pi |g_0|^2  \int d\eps'\, \rho_{\rm e}(\eps') \int\, d\Omega \; \rho_{\rm ph}(\Omega)  \nonumber \\	 &\times \bigg[ \big( n_B(\Omega) +1 - f(\eps' )\big) \delta(\eps-\eps'  -\hbar\Omega) \nonumber \\
 & +\big(  n_B(\Omega) +f(\eps' ) \big) \delta(\eps-\eps' +\hbar\Omega)	\bigg],
\end{align}
where $|g_0|^2$ is the coupling strength, $T$ is the temperature, $n_B(\Omega)$ is the Bose distribution, and $\rho_{\rm ph}(\Omega)$ is the phonon density of states, which we take to be a Lorentzian centered around $\hbar\Omega_0$ with a phenomenological broadening $\Delta$. Finally, $\rho_{\rm e}(\eps)$ is the electron density of states, which can be determined through the recursive algorithm described above as $\rho_{\rm e} (\eps) = -(1/\pi)\sum_{\rrr} \im\{ G^r(\rrr, \rrr,\eps)\}$. In the low-temperature limit, there is no phonon annihilation and the expression reduces to
\begin{align}
&\im  \{\Sigma_{\rm ph}(\eps,T\rightarrow 0)\} =-\pi |g_0|^2 \int d\Omega \; \rho_{\rm ph}(\Omega) \nonumber \\
	 &\times \bigg[  \big( 1-f(\eps-\hbar \Omega)\big)\rho_{\rm e}(\eps-\hbar\Omega)  \nonumber \\
 & +f(\eps +\hbar\Omega\big)\rho_{\rm e}(\eps+\hbar\Omega)    \bigg].\eqlab{sigma_ph}
\end{align}
The real part of the self-energy corresponds to the energy shift induced by the interactions and can be conveniently determined using the Kramers--Kronig relation (see details in the Appendix).

 \begin{figure}[htb!]
 	\begin{center}
 		\includegraphics[width= 0.95\columnwidth]{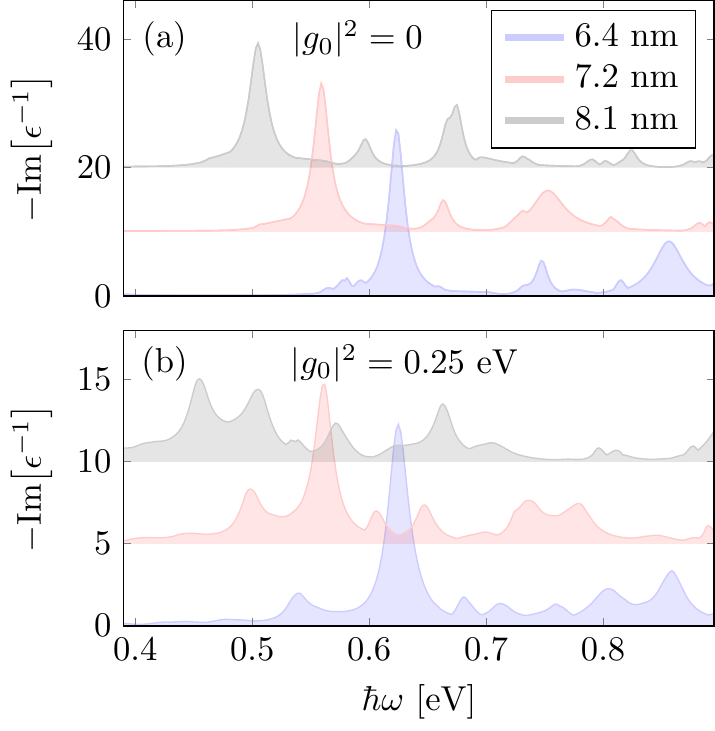}
 		\caption[]{Eigenvalue loss-spectrum for different sizes of armchair graphene nanotriangles. Panel (a) uses an energy-independent broadening $\eta = 5$\,meV whereas panel (b) includes the phonon interaction with an electron-phonon coupling constant $|g_0|^2=0.25$\,eV\cite{Park2007,Jablan2009_PRB} and the optical phonon centered at $\hbar\Omega_{0}=0.2$\,eV with broadening $\Delta=5$\,meV.} \label{fig:multi_eels}
 	\end{center}
 \end{figure}

In \figref{multi_eels}, we show the eigenvalue loss-spectra of nanotriangles of different side lengths with and without inclusion of the phonon interaction, as described by \eqref{sigma_ph}. To increase visibility, we have neglected the redshift caused by the real part of the self-energy and aligned the position of the high energy dipole peak to the one without the phonon coupling.

When examining \figref{multi_eels}(a) and \ref{fig:multi_eels}(b), we find a well-known blueshift for decreasing structure size\cite{Christensen:2014,Yan:2013}.
Comparing the two panels, we observe that phonon interaction induces additional peaks in the spectrum, which are caused by hybridization between the plasmon and phonon modes. In particular, the dipole mode exhibits a strong hybridization and splits into two distinct peaks. This peak splitting displays the characteristic behavior of an avoided crossing mechanism\cite{Hwang2010_PRB,Jablan:2011} between the phonon and plasmon mode, as also observed experimentally for graphene nanoribbons\cite{Yan:2013} and graphene nano-disks.\cite{Zhu:2014}
Although the spectral width remains almost constant as the size changes, the spectral weight is transferred between the two hybridized dipole peaks, as revealed by comparing the spectra for different sizes in \figref{multi_eels}(b).

   \begin{figure}[htb!]
 	\begin{center}
 		\includegraphics[width= 0.95\columnwidth]{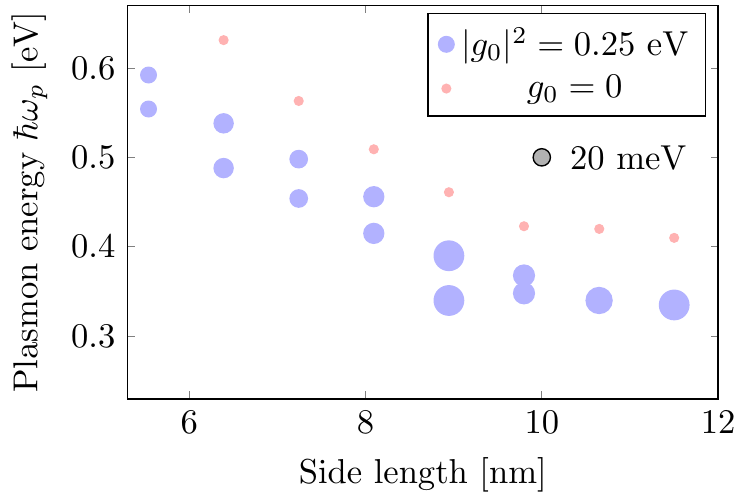}
 		\caption[]{Spectral position of the dipole plasmon peak as a function of the side length of the armchair graphene nanotriangle. The radius of the marks indicate the full-width-half-maximum of the plasmon peak for a phonon density broadening of $\Delta=5$\,meV.} \label{fig:peak_vs_size}
 	\end{center}
 \end{figure}

The existence of peak splitting signals a strong coupling that exceeds the intrinsic plasmon linewidth. For example, in the dipole peak near 0.5\,eV for 8.1\,nm structures in \figref{multi_eels}(a), we notice that a strong coupling regime is reached in which the plasmon-phonon hybridization leads to two dipole peaks below 0.5\,eV for 8.1\,nm structures in \figref{multi_eels}(b), as opposed to the single peak in the absence of phonons in \figref{multi_eels}(a). These features are represented in \figref{peak_vs_size}, where we observe similar effects consistently over a large range of triangle sizes. Evidently, this strong coupling leading to peak splitting is more effective for small nanotriangles thereby making phonon interactions especially interesting in the regime of molecular 2D plasmons.
In the limit of large structures the plasmon-phonon hybridization becomes less pronounced (results not shown), and we therefore return to the weak coupling regime, in which the phonon interaction broadens and redshift the plasmon peak.

Plasmon broadening is indicated on \figref{peak_vs_size} through the size of the data points (i.e., the radius indicates the full-width-half-maximum linewidth of the resonance). For most cases the linewidth of individual peaks are below $\sim$30\,meV, corresponding to plasmon lifetimes of $\sim$20\,fs. This is a few times larger than the lifetimes of localized surface plasmons in noble metal nanostructures,\cite{Maier:2007} but still shorter than the expected lifetimes for plasmons in high-quality extended graphene samples.\cite{Woessner:2015}

\begin{figure}[htb!]
 	\begin{center}
 		\includegraphics[width= 0.95\columnwidth]{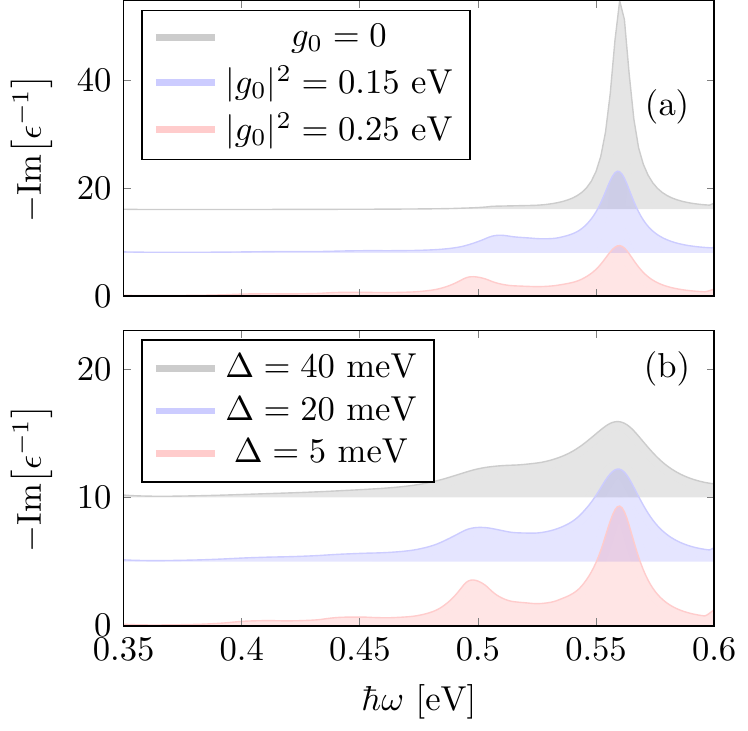}
 		\caption[]{(a) Dipole peak for different values of the electron-phonon coupling in the armchair nanotriangle with side length of 7.2\,nm  ($\eta = 10$\,meV). (b) The same dipole peak as in (a) for $|g_0|^2 = 0.25$\,eV, but for different values of the broadening, $\Delta$, of the phonon density of states. } \label{fig:dipole_g_delta}
 	\end{center}
 \end{figure}

As a final illustration of plasmon-phonon coupling, we vary both the electron-phonon coupling strength and the broadening of the phonon density of states, as shown in \figref{dipole_g_delta}. The peak splitting increases with phonon coupling (\figref{dipole_g_delta}(a)), while we require a narrow phonon mode to obtain a well-defined hybridization and therefore distinguishable hybridization peaks. We insist once more that these calculations also apply for extrinsic phonons of a substrate, incorporated exactly through the same formalism, with appropriately chosen coupling strength and phonon frequency.

\section{Discussion}
In this work, we have included a self energy $\Sigma(\eps)$ in the single-electron Green function which accounts for the interactions between electrons and other degrees of freedom,\cite{Haug:2008} and in particular, phonons are considered in the presented examples. Naturally, the self-energy could also account for the life-time broadening of electron states in open quantum systems,\cite{Datta:1995} where electrons, and consequently plasmons, are quasi-localized to regions of finite extension within an otherwise effectively bulk sheet of graphene. The list of potential geometries can include narrow constrictions (quantum point contacts) in an otherwise infinite ribbon or sheet of graphene,\cite{Tombros:2011} as well as finite graphene anti-dot arrays\cite{Pedersen:2008,Jin:2017,Pan:2017} in extended graphene. We also include on this list non-planar regions in an otherwise planar sheet of graphene with a local finite surface curvature acting as a trapping potential for electrons and plasmons.\cite{Wang:2016,Smirnova:2016,Goncalves:2016a,Goncalves:2017,Fei:2016,Slipchenko:2017} Localization could likewise be due to local strain and pseudo-magnetic fields.\cite{Settnes:2016} The explicit energy dependence of $\Sigma(\eps)$ is common to these problems and reflects the energy dependence of the density-of-states in the surrounding medium. Because of this, the life-time broadening would not be captured accurately by a phenomenological constant damping rate. 

\section{Conclusion}

We have introduced a computationally highly efficient approach to describe the optical response of plasmonic nanostructures that facilitates the account for many-body interactions. Our approach neglects vertex corrections, which is a valid approximation for metallic nanostructures or 2D Dirac-fermion systems, such as graphene. Here, we have illustrated the power of our method for armchair graphene nanotriangles of various sizes, paying special attention to the evaluation of eigenvalue loss-spectra for plasmons dressed by optical phonons. The interactions with phonons are represented by self-energies that enter the electron Green function, and we emphasize that the energy dependence of the self-energies leads to qualitative changes in the spectra that cannot immediately be accounted for by a more phenomenological broadening, such as inclusion of a complex-valued substrate dielectric function. As an example, the hybridization between plasmons and phonons manifests in dramatic peak splitting. While we have focused on interactions with phonons, we may without further complications apply our self-energy formalism to account for other types of nanostructures and interactions relevant to quantum plasmonics,\cite{Tame:2013,Fitzgerald:2016,Bozhevolnyi:2017a,Zhu:2016a} including also dephasing phenomena and life-time broadening in open quantum systems, where the electron gas remains only quasi confined.

\begin{acknowledgement}
\noindent
{\footnotesize
We thank Christian Wolff for stimulating discussions on the numerics and useful comments on an initial version of this manuscript.
The Center for Nanostructured Graphene is supported by the Danish National Research Foundation (DNRF103).
This work is partially supported by the European Commission (Graphene Flagship CNECT-
ICT-604391 and FP7-ICT-2013-613024-GRASP) and the Spanish MINECO (MAT2014-59096-P , Fundaci\'o Privada Cellex, and SEV2015-0522). J.~R.~M. Saavedra acknowledges financial
support through AGAUR FI B 00492-2015.
M.~S. and N.~A.~M. acknowledge support from the Danish Council for Independent Research (DFF-5051-00011 \& DFF 1323-00087). 
N.~A.~M. is a VILLUM Investigator supported by VILLUM Fonden.
}

\end{acknowledgement}

\section{Appendix: Kramers--Kronig relation}

We determine the real part of $\chi^0_R(\omega)$ from the imaginary part $\chi^0_I(\omega)$. Due to causality of the response function, this can be done using the Kramers--Kronig relation
 \begin{equation}
  \chi^0_R(\omega)=\frac{1}{\pi}P\int_{0}^{\infty}dx\, \frac{ \chi^0_I(x)}{x - \omega} +
  \frac{1}{\pi}P\int_{0}^{\infty}dx\, \frac{ \chi^0_I(x)}{x + \omega},
   \end{equation}
where $P$ denotes Cauchy's principal part. The Kramers--Kronig integration can be replaced by a weighted sum,\cite{Shishkin2006_PRB}
\begin{subequations}
\begin{equation}
\chi^0_R(\omega_i) = \sum_n W_n(\omega_i) \chi^0_I(\omega_n),
\end{equation}
where the weight factors are calculated using
\begin{equation}
W_n(\omega_i)  = \frac{1}{\pi}P\int_0^{\infty} dx\,\Phi_n(x) \bigg(\frac{1}{x - \omega_i}+\frac{1}{x + \omega_i}\bigg)
\eqlab{wfactor}
\end{equation}
with

\begin{equation}
\eqlab{phi_n}
\Phi_n(x) = \left\{\begin{matrix}\frac{\omega_{n+1}-x}{\omega_{n+1}-\omega_n}&,& \omega_n \leq x \leq \omega_{n+1}\\
\frac{x-\omega_{n-1}}{\omega_{n}-\omega_{n-1}}&,& \omega_{n-1}  \leq x \leq \omega_{n}\\0&,&{\rm otherwise}
\end{matrix} \right.
\end{equation}

%\begin{equation}\eqlab{phi_n}
%\Phi_n(x) = \frac{\omega_{n+1}-x}{\omega_{n+1}-\omega_n}, \quad \quad \omega_n \leq x \leq \omega_{n+1}
%\end{equation}
%\begin{equation}
%\Phi_n(x) = \frac{x-\omega_{n-1}}{\omega_{n}-\omega_{n-1}}, \quad \quad \omega_{n-1}  \leq x \leq \omega_{n}
%\end{equation}
%\begin{equation}
%\Phi_n(x) = 0 \quad \quad {\rm otherwise}
%\end{equation}
\end{subequations}
To determine the weight factors, we consider the integral \eqref{wfactor} and insert \eqref{phi_n} to obtain an analytical expression for $W_n$,
\begin{multline}
W_n(\omega_i) = \frac{1}{ \pi (\omega_{n }-\omega_{n-1}) } \bigg[ 
( \omega_i -\omega_{n-1})   \log\bigg|\frac{\omega_n-\omega_i}{\omega_{n-1}-\omega_i} \bigg| \\ 
-(\omega_i +\omega_{n-1})\log\bigg|\frac{\omega_n+\omega_i}{\omega_{n-1}+\omega_i} \bigg| \\- ( \omega_i - \omega_{n+1}) \log\bigg|\frac{\omega_{n+1}-\omega_i}{\omega_{n}-\omega_i} \bigg|\\+ (\omega_i +\omega_{n+1}) \log\bigg|\frac{\omega_{n+1}+\omega_i}{\omega_{n}+\omega_i} \bigg|
\bigg].
\end{multline}
This procedure allows us to replace the integral with a summation in order to obtain $\chi_R^0$ on a grid containing the mid-points of the original grid in which $\chi^0_I$ was determined. Linear interpolation can be used efficiently to produce $\chi^0_I$ and $\chi^0_R$ on the same grid.

\bibliography{graphene}

\end{document}